\documentstyle[epsfig,longtable]{aipproc}

\begin{document}
\title{SNEWS: A Neutrino Early Warning System for Galactic SN II}

\author{Alec Habig$^*$ for the SNEWS collaboration} 

\address{$^*$Boston University Physics Dept., Boston, MA 02215}
  

\maketitle

\begin{abstract}
  The detection of neutrinos from SN1987A confirmed the core-collapse
  nature of SN II, but the neutrinos were not noticed until after the
  optical discovery.  The current generation of neutrino experiments are
  both much larger and actively looking for SN neutrinos in real time.
  Since neutrinos escape a new SN promptly while the first photons are
  not produced until the photospheric shock breakout hours later, these
  experiments can provide an early warning of a coming galactic SN II.
  A coincidence network between neutrino experiments has been
  established to minimize response time, eliminate experimental false
  alarms, and possibly provide some pointing to the impending event from
  neutrino wave-front timing.
\end{abstract}

\section*{Introduction}
In a supernova (SN) driven by the gravitational collapse of a massive
stellar core into a neutron star ({\it e.g.} Type II or Type Ib SNe),
the resulting huge pulse of neutrinos escapes the star promptly after
the collapse (for a detailed review and discussion of SN neutrinos, see
\cite{Burrows:1992kf}).  In the case of the nearby SN1987A, these
neutrinos were observed by the Kamiokande\cite{Hirata:1986hu} and
IMB\cite{Bionta:1987qt} neutrino detectors in offline analyses performed
after the optical discovery of the SN.  Although the neutrinos escape
the star promptly, photons do not get out until the shock wave travels
from the core through the stellar envelope and breaks out of the stellar
photosphere -- hours later, depending upon the size of the envelope
(ref.~\cite{lagtime} contains a simple model of this delay).  Thus, if
the neutrinos could be detected in real-time, they would provide advance
warning of the coming photons from the SN.  Such a warning would allow
preparation of observations over the whole electromagnetic spectrum,
maximizing the information gathered during the previously unobserved
first few hours of light from a new SN.
  
The {\underline S}uper{\underline n}ova {\underline E}arly {\underline
  W}arning {\underline S}ystem (SNEWS)~\cite{Scholberg:1999tm} is a
coincidence trigger between the world's neutrino telescopes.  While the
individual neutrino detectors all have near real-time SN monitors which
are sensitive to SNe in the Milky Way galaxy and its satellites ({\it
  e.g.}~\cite{Aglietta:1992dy,Oyama:1994zq,Ambrosio:1997hh}), those
monitors could be fooled by instrumental effects.  To avoid releasing
such a false alarm, any single experiment would want its SN monitor
output to be carefully scrutinized by a human.  Unfortunately, the time
scale on which humans operate is of the same order as the lead time
gained by the neutrinos over the photons, which would waste much of the
advance notice.  Those extra hours would be better spent by observers
preparing to observe the impending photons.  However, the likelihood of
two independent experiments experiencing a false alarm in coincidence is
very small, therefore an automated alert can be issued with confidence.
If each input experiment has a false alarm rate of $<1/$week, the false
coincidence rate will be $\ll1/$century.  Thus, the SNEWS network can
eliminate the need for active human supervision and provide an alert to
the astronomical community as promptly as possible by providing an
automated alarm when multiple experiments simultaneously see a SN-like
neutrino signal.

\section*{The SNEWS network}

There are a number of experiments capable of detecting neutrinos from a
galactic SN (Tab.~\ref{tab:nuexp}) either active now or coming online
within the next few years.  Experiments participating in the SNEWS
network send a standard UDP packet via the internet to a remote server
when their local automated SN monitor detects a SN-like signal in their
detector.  This remote server forms a blind coincidence trigger between
incoming alert packets and issues an alarm to interested observers
should it record multiple experimental SN triggers time-stamped within
10 seconds (in UTC) of each other.  

\begin{table}
\begin{tabular}{||c|c|c|c|c|c||}\hline\hline
Detector & Type & Mass & Location  & \# events  & Status\\ 
         &      & (kton) &         & @10 kpc   & \\\hline
{\bf Super-K} & water  & 32    & Japan  & 4400 & {\bf signaling SNEWS} \\ 
        & Cherenkov  & &        & &  {\bf since May 1998}  \\ \hline
{\bf MACRO} & scint.& 0.6 & Italy & 150 &  {\bf signaling SNEWS} \\
             &       &     &       & &   {\bf since March 1998}\\ \hline
{\bf LVD} & scint. & 0.7 & Italy & 170 & {\bf signaling SNEWS}\\ 
             &       &     &       & &   {\bf since Feb. 1999}\\ \hline
SNO & H$_2$O,& 1.7 & Canada &  350 & running \\
    & D$_2$O & 1   &        &  430 &  \\ \hline
AMANDA & long   & M$_{\rm eff}\sim$ & Antarctica  & N/A & running \\ 
       & string &2/pmt &        &  &        \\ \hline
Baksan & scint. & 0.33  & Russia  & 70 &  running\\ \hline
Borexino & scint. & 1.3 & Italy & $\sim$200 & 2000       \\\hline 
Kamland & scint. & 1 & Japan & 300 & 2001       \\\hline 
OMNIS & high Z   & 10~kT~Fe, &  USA & 2000 &  2000+\\ 
      &          & 4~kT-Pb  &      &      &  \\\hline 
LAND    & high Z & 1 & Canada & 450 & 2000+\\\hline
Icanoe & liquid argon & 9 & Italy & & 2000+ \\ \hline\hline
\end{tabular}
\caption{Current and near-future neutrino detectors capable of detecting
  the neutrino signal from a galactic core-collapse SN. {\bf Bold}
  entries are currently providing input to the SNEWS network.}
\label{tab:nuexp}
\end{table}

When the SNEWS server sees such a coincidence between experiments, it
will issue an alert via email to all interested parties.  Email to pager
gateways will provide the fastest initial notification.  Currently the
MACRO, Super-Kamiokande, and LVD experiments are providing automated
inputs to this coincidence trigger.  It is anticipated that the SNO and
AMANDA experiments will begin providing inputs early in the year 2000.
To verify the prediction of a vanishingly small rate of false alarms,
the coincidence server is presently running in a test mode that would
not send out an automated alarm.  Since no false coincidences have
occurred, SNEWS will be switched over to its final, fully automated
configuration in the year 2000.

\section*{What Information can SNEWS Provide?}

The single most important information provided by SNEWS would be simply
the fact that within hours, a nearby SN will soon be visible.  If
nothing else, this alarm will allow observers get set up so as to be
able to start observing as soon as more information becomes available.

Naturally more information would be desirable, particularly that which
tells observers where to look.  There are two methods which could
provide pointing information from the neutrino signal.  The first takes
advantage of the directionality of some neutrino interaction channels in
individual detectors, most notably electron scattering
($\nu_x~+~e^-~\rightarrow~\nu_x~+~e^-$) in the large water Cherenkov
detectors.  For a SN at 10~kpc, it is estimated that Super-K could point
to a $\sim5^\circ$ cone on the sky, and SNO a $\sim20^\circ$
cone~\cite{Beacom:1998fj}.  While hardly precise by photon astronomy
standards, these solid angles are easily covered by large field of view
instruments.  This directional information is not a product of the SNEWS
network but that of the individual experiments' analyses, but SNEWS
could play an important role in disseminating and correlating such
information as it becomes available.

The second pointing method is unique to SNEWS.  The precise timing of
the neutrino wavefront arrival at the different detectors' widely
separated locations on the earth could be used to reconstruct where the
neutrinos were coming from.  Unfortunately, a detailed
analysis~\cite{Beacom:1998fj} of the statistics available to the current
detectors suggests that this ``triangulation'' approach would be
substantially less precise than the $\nu+e$ scattering, being mostly
valuable as a confirmation rather than as a position refinement -- but
this itself is both a helpful and important cross-check.

\section*{Plans for use of an Early Warning}

Should a core-collapse SN occur in the range of the neutrino detectors
(within the Milky Way galaxy or its satellites), an alert would be issued
hours before light is produced by the new SN.  This short detection
range ensures that any SN signal triggering SNEWS will be very nearby
and thus a potential scientific bonanza.  Conversely, this short range
translates to a probe of a very limited volume, making for a low
observable SN rate~\cite{snrate} of the order of several SNe per century.
In order to make the most of this once in a lifetime opportunity, good
plans should be developed in advance, to be put into action when a SN
does occur.

Two examples of such planning are currently in place.  The first is from
the editors of {\it Sky \& Telescope} magazine, who are preparing to
both provide a means to disseminate the alarm information to the amateur
astronomer community, and to manage the resulting flow of observations
such an alarm would generate~\cite{skyandt}.  The celebrated accidental
observations of SN1987A by the amateur astronomer Albert Jones were
instrumental in studying the early stages of this important event.  An
organized effort by the many similarly skilled observers available would
not only provide many more valuable early observations, but also aid
greatly in precisely locating the new SN, given both the wide
angle instruments commonly used by amateurs and their sheer numbers,
experience with the sky, and enthusiasm.

The second example is a Hubble Space Telescope Target of Opportunity
proposal from J.~Bahcall {\it et al}~\cite{hst-too}.  Given the long
response time of the HST and the precise positional information needed to 
point it, these observations would not occur until days after the
initial alarm.  However, the HST is the only facility available to take
the high resolution UV spectra needed to understand the nearby
environment of the new SN, and having the detailed observation plans on
file and ready to be used would not only save time but provide better
results when they are needed.

Any astronomer who might want to observe such an important event is
encouraged to make similar plans.  Plans and preparations now not only
will make good observations easier when the exciting but hectic time
comes to observe a galactic SN, but there are undoubtably many good
ideas as yet unborn.  Some good thinking now in the idle years while
waiting for such an event will allow these ideas to be fleshed out and
properly put into practice when it counts.

%

\end{document}